\documentclass[conference]{IEEEtran}
\usepackage{pifont}
\usepackage{amsfonts}
\usepackage{subfigure}

\usepackage{amssymb}
\usepackage{amsmath}
\usepackage{epsfig}
\usepackage{color}
\usepackage{url}

\usepackage{mathabx} 

\begin{document}
\newtheorem{theorem}{Theorem}
\newtheorem{corollary}{Corollary}
\newtheorem{conjecture}{Conjecture}
\newtheorem{definition}{Definition}
\newtheorem{lemma}{Lemma}
\newtheorem{algorithm}{Algorithm}
\newtheorem{remark}{Remark}
\newtheorem{idea}{Idea}
\newtheorem{observation}{Observation}

\newcommand*{\QEDB}{\hfill\ensuremath{\square}}%
\newcommand{\define}{\stackrel{\triangle}{=}}

\def\argmax{\mathop{\rm arg\,max}}
\def\argmin{\mathop{\rm arg\,min}}
\def\minimize{\mathop{\rm minimize}}

\newcommand{\thetav}{\hbox{\boldmath$\theta$}}
\newcommand{\Rmh}{\widehat{{\bf R}}}

\pagestyle{empty}

\def\sDoF{\overline{\mbox{\normalfont \scriptsize DoF}}}

\def\QED{\mbox{\rule[0pt]{1.5ex}{1.5ex}}}
\def\proof{\noindent{\it Proof: }}
\twocolumn
\date{}

\title{On-the-fly Large-scale Channel-Gain Estimation for Massive Antenna-Array Base Stations}
\author{\IEEEauthorblockN{Chenwei Wang$^*$, Ozgun Y. Bursalioglu$^*$, Haralabos Papadopoulos$^*$, Giuseppe Caire$^{\dagger}$}\\
\IEEEauthorblockA{$^*$DOCOMO Innovations, Inc., Palo Alto, CA\\
$^{\dagger}$Technische Universit¨at Berlin, Germany\\
$^*$\{cwang, obursalioglu, hpapadopoulos\}@docomoinnovations.com,\, $^{\dagger}$caire@tu-berlin.de}}
\maketitle

\thispagestyle{empty}
\begin{abstract}
We propose a novel scheme for estimating the large-scale gains of the channels between user terminals (UTs) and base stations (BSs) in a cellular system. The scheme leverages TDD operation, uplink (UL) training by means of properly designed non-orthogonal pilot codes, and massive antenna arrays at the BSs. Subject to $Q$ resource elements allocated for UL training and using the new scheme, a BS is able to estimate the large-scale channel gains of $K$ users transmitting UL pilots in its cell and in nearby cells, provided $K\leq Q^2$. Such knowledge of  the large-scale channel gains of nearby out-of-cells users can be exploited at the BS to mitigate interference to the out-of-cell users  that experience the highest levels of interference from the BS. We investigate the large-scale gain estimation performance provided by a variety of non-orthogonal pilot codebook designs. Our simulations suggest that among all the code designs considered, Grassmannian line-packing type codes yield the best large-scale channel gain estimation performance.

\end{abstract}

\allowdisplaybreaks

\section{Introduction}\label{sec:intro}

Massive MIMO, originally introduced by Marzetta in \cite{Marzetta_massive_mimo}, has been widely recognized as a key enabling technology of 5G/B5G. To harvest massive MIMO gains in the downlink (DL) with multiuser (MU)-MIMO, channel state information (CSI) of the users is needed at the BS, referred to as CSI acquisition. In the FDD systems, UTs first learn the channels based on DL reference signals and subsequently feed their CSI to BS in the UL. The overwhelming CSI-overhead trends of conventional FDD-based schemes, such as those in FDD-based 4G-LTE systems, have spurred a lot of activities and more efficient schemes design \cite{Ansuman-JSDM}. Alternatively, in TDD systems, DL CSI can be directly obtained at the BS from UL training  and by capitalizing on the principle of UL/DL channel reciprocity \cite{Marzetta_massive_mimo}. Reciprocity-based schemes offer fast  acquisition of channels between users and a massive BS array and with low overheads, as they require allocating as few UL pilot dimensions as the number of simultaneously served single-antenna users \cite{Marzetta_noncooperative}.

At higher carrier frequency (e.g., mmWave) bands where large chunks of bandwidth would be available, more massive antennas can be readily deployed on the same footprint due to the shorter wavelengths. Also, due to physics, channels decorrelate much faster at these bands than at the sub-6GHz bands used by LTE, which make the fast low-overhead CSI acquisition of TDD/reciprocity-based schemes even more attractive. In addition, at these higher frequencies, dynamic shadowing and intermittent signal blocking due to the appearance of obstacles lead to shorter intermittent coverage and dramatic pathloss swings, implying the need for much denser, inherently irregular BS/radio-head deployments.

In densely deployed inherently irregular networks, however, managing resource allocation and interference becomes very challenging. For example, the notion of traditional cell planning (along with traditional frequency reuse) is no longer practical, and new access techniques are needed (whether slotted or random access) that readily scale with the densification of the infrastructure. Recently, a class of non-orthogonal pilot codes were proposed in \cite{Ozgun_ICC2016} for radio remote head (RRH) networks, called ``Spotlight", where users using common pilot dimensions over a slotted-access system are opportunistically served by a subset of nearby RRHs. Spotlight relies on aggressive pilot reuse where many users are aligned on a UL pilot dimension, and exploits  very fast user detection at each RRH based on a simple binary energy detection scheme. Although \cite{Ozgun_ICC2016} focuses on slotted scheduled access, the same principles can be employed in the context of random access.

In this paper, we focus on a TDD/OFDM-based system which relies on reciprocity-based CSI acquisition, much like the ones considered in e.g., \cite{Marzetta_massive_mimo} and \cite{Ozgun_ICC2016}. Assuming slotted transmission, we consider the OFDM plane is split into resource blocks (RBs), each of which consists of resource elements (REs) within the coherence time and bandwidth of the channel. Also, assuming $Q$ REs in an RB are allocated for UL training, we consider the scenario where $K$ users are in the vicinity of a massive-array BS and transmit pilots over these $Q$ pilot REs using pre-assigned codewords. Focusing on a quasi-static channel model where the user-channels stay constant within any given RB, we study the problem of estimation of the large-scale gains of the user channels based on massive-array observations over the $Q$ pilot REs. In particular, we first determine the conditions on the number of users and the number of pilot REs which ensure large-scale channel gain estimation is feasible. Then we develop and analyze high-performing pilot designs. Due to the space limitation, for more design options and analysis, including extensions to frequency-selective MIMO channel models, please see the extended version \cite{wbe_full}.

Designing the UL pilot frame to enable the BS to learn large-scale channel gains in addition to small-scale channels has several important application scenarios. For instance, it can be used in the context of dense small-cell deployments to allow a BS to learn the large-scale channel gains of the strong (nearby) out-of-cell users simultaneously served by nearby BSs, so that it can subsequently exploit them for interference mitigation purposes, which we will investigate in detail in \cite{wbe_full}. Another example where the need to learn these large-scale channel gains naturally arises, involves slotted random access, where the BS would also wish to know the subset of the $K$ users that decided to access the medium during any given RB.

\section{System Model}\label{sec:system}

We consider a setting where a single BS is equipped with an $M$-element antenna array and $K$ single-antenna UTs are randomly distributed around the BS (e.g., in its cell and in nearby cells).  We assume TDD operation over OFDM, and employ a quasi-static channel model where each user channel remains constant within an RB comprising a set of $T$ OFDM REs within the user-channel coherence time and bandwidth.

To provide context for the problem of interest, we consider reciprocity-based training for DL transmission over a generic RB, and the DL user channels  are learned  at the BS via  UL user pilots broadcasted within that RB\footnote{Due to  to radio channel reciprocity, the UL and DL radio channels between the BS and a given user are the same within  any given RB.}. Moreover, we assume that $Q$ REs are allocated for UL pilot transmission in the RB, and user $k$ broadcasts a pre-assigned $Q\times 1$ pilot pattern, denoted by ${\bf p}_k$, over these $Q$ REs. When $K$ users simultaneously transmit their pilot patterns over an RB to train their channels, the received signal at the BS is given by:
\begin{subequations}
\begin{equation}
{\bf Y}=\sum_{k=1}^K {\bf p}_k{\bf h}_k^T+{\bf W}
\label{eqn:sys_model}
\end{equation}
where
\begin{equation}
{\bf Y}=
\begin{bmatrix}
{\bf y}_{(1)} \! & \!\cdots\! & \!{\bf y}_{(M)}
\end{bmatrix}  \text{ and  }
{\bf W}=
\begin{bmatrix}
{\bf w}_{(1)} \!&\! \cdots \!& \! {\bf w}_{(M)}
\end{bmatrix}
\end{equation}
\end{subequations}
are both $Q\times M$ matrices. The $M\times 1$ random vector ${\bf h}_k=[h_{k1}\cdots h_{kM}]^T \sim \mathcal{CN}({\bf 0},g_k{\bf I}_M)$ represents the UL/DL channel between the BS and user $k$ within the RB, with $g_k$ representing the large-scale gain of the user $k$'s channel. In addition, the $Q\times 1$ vectors ${\bf y}_{(m)}$, ${\bf w}_{(m)}\sim \mathcal{CN}({\bf 0},\sigma_w^2{\bf I}_M)$ represent the received signal vector and the additive white Gaussian noise vector at the $m^{th}$ antenna over the $Q$ REs respectively, and they satisfy
\begin{eqnarray}
{\bf y}_{(m)}=\sum_{k=1}^K  {\bf p}_k h_{km}+{\bf w}_{(m)}={\bf P}{\bf h}_{(m)}+{\bf w}_{(m)}\label{eqn:signal_m}
\end{eqnarray}
where ${\bf h}_{(m)}\triangleq [h_{1m}\cdots h_{Km}]^T$ is a $K\times 1$ vector and ${\bf P}\triangleq[{\bf p}_1~\cdots~{\bf p}_K]=(p_{qk})_{Q\times K}$ is a $Q\times K$ matrix.

In this paper, we focus on the problem of estimating the large-scale user-channel gains $\{g_k\}$ based on the observation of ${\bf Y}$ in (\ref{eqn:sys_model}) and {\em a priori} knowledge of the pilot sequences ${\bf P}$ pre-assigned to the BS.

Instances of such a problem where the large-scale gains are unknown and need to be estimated naturally arise in cellular networks, whereby each BS independently serves UTs in its cell. In the context of slotted transmission, the $K$ nearby users transmitting UL pilots also include all scheduled users in neighboring cells whose channel gains are unknown at the given BS. Learning the out-of-cell user's channel gains can be used for combating interference to the nearby (strong) out-of-cell users. The problem also naturally arises in the context of slotted random-access, where $K$ represents the number of users in the system, and only a fraction of them are active (i.e.,  access the UL channel with pilot transmissions) within a given RB. In this case, $g_k$ represents the ``effective'' channel gain of user $k$ (which is zero if user $k$ is not active) and can be thus treated as unknown at the BS. In either case, it is natural to estimate the large-scale channel gains of the $K$ users first, and then proceed to estimation of the small-scale channel coefficients of the dominant/detected users.

In the following, we provide our solutions to the large-scale gain estimation problem in Sec. \ref{sec:scheme}. In the process we also determine the maximum number of users, $K$, that can be supported for a given number of pilot REs, $Q$, in the sense that all $K$ large-scale gains can be estimated via the observation in (\ref{eqn:sys_model}). In Sec. \ref{sec:performance} and  \ref{sec:sim}, we present theoretical and simulation-based performance analysis for a class of pilot designs.

\section{Proposed Scheme: Extracting Large-Scale Channel Gains with Large Arrays}\label{sec:scheme}\label{sec:scheme_large}

Our proposed method estimates the large-scale channel gains $\{g_k\}$ using the sample covariance matrix of ${\bf Y}$:
\begin{equation}
\label{eqn:Ryh}
\Rmh_{\bf y}\triangleq \frac{1}{M}{\bf Y}{\bf Y}^H=  \frac{1}{M}\sum_{m=1}^M {\bf y}_{(m)}{\bf y}_{(m)}^H.
\end{equation}
In particular, the method leverages the presence of a large array (i.e., large $M$) at the BS, where the sample covariance $\Rmh_{\bf y}$ in (\ref{eqn:Ryh}) converges to the covariance of  ${\bf y}_{(m)}$. The next theorem gives conditions under which (accurate) estimation of the $\{g_k\}$ is possible with large arrays at the BS.

{\bf Theorem 1:} Consider the matrix ${\bf P}=(p_{qk})_{Q\times K}$ where $p_{qk}$ is complex and generic. When $M\rightarrow \infty$, the BS can estimate the large-scale channel gains $g_k$'s of all the $K$ users from (\ref{eqn:Ryh}) almost surely if $K\leq Q^2$.

\emph{Proof:}
Since ${\bf h}_{(m)}$ and ${\bf w}_{(m)}$ are both i.i.d. over $m$, (\ref{eqn:signal_m}) automatically implies that ${\bf y}_{(m)}$ is also i.i.d. over $m$. Hence, we can drop the foot script $m$ of the relevant notations for brevity. Also, when $M\rightarrow \infty$, $\Rmh_{\bf y}$ converges to its mean $\mathbb{E}[\Rmh_{\bf y}]$ almost surely, where
\begin{eqnarray}
\mathbb{E}[\Rmh_{\bf y}]=\mathbb{E}[{\bf y}_{(m)}{\bf y}_{(m)}^H].
\end{eqnarray}
Letting ${\bf R}_{{\bf y}}$ denote the covariance of ${\bf y}_{(m)}$, we have
\begin{eqnarray}
\lim_{M\rightarrow \infty}\frac{1}{M^2}\|\Rmh_{\bf y}-{\bf R}_{\bf y}
\|_{\textrm{F}}^2= 0,
\end{eqnarray}
which implies that, if $M\rightarrow \infty$, then ${\bf R}_{\bf y}$ is also available at the BS almost surely.

Next, we elaborate on how to estimate $\{g_k\}$ from ${\bf R}_{\bf y}$. First of all, we decompose ${\bf R}_{\bf y}$ in the following form:
\begin{eqnarray}
{\bf R}_{\bf y}&\!\!\!\!=\!\!\!\!& {\bf P}\mathbb{E}[{\bf h}_{(m)}{\bf h}_{(m)}^H]{\bf P}^H+\mathbb{E}[{\bf w}_{(m)}{\bf w}_{(m)}^H]\\
&\!\!\!\!=\!\!\!\!&{\bf P}{\bf G}{\bf P}^H+\sigma_w^2{\bf I}\label{eqn:theo_cov}
\end{eqnarray}
where ${\bf G}=\textrm{diag}([g_1,g_2,~\cdots~ g_K])$.
Rewriting (\ref{eqn:theo_cov}) as
\begin{equation}
{\bf R}_{\bf y}-\sigma_w^2{\bf I}=\sum_{k=1}^Kg_k{\bf p}_k{\bf p}_k^H\label{eqn:theo_cov2}
\end{equation}
yields $Q^2$ linear equations with the $g_k$'s as the only unknowns. Alternatively, (\ref{eqn:theo_cov2}) is a system of $Q^2$ linear equations and $K$ unknowns. To see this, we rewrite (\ref{eqn:theo_cov2}) into a canonical form. Specifically, we first reshape ${\bf R}_{\bf y}$ and $\sigma_w^2{\bf I}$ to
\begin{eqnarray}
{\bf r}_y&\!\!\!\!\triangleq\!\!\!\!& \textrm{vec}({\bf R}_{\bf y})=[{\bf r}_{y1}^T~\cdots~{\bf r}_{yQ}^T~]^T,\label{eqn:reshapeR}\\
{\bf r}_w&\!\!\!\!\triangleq\!\!\!\!& \textrm{vec}(\sigma_w^2{\bf I})=\sigma_w^2[{\bf e}_1^T~\cdots~{\bf e}_Q^T~]^T,
\end{eqnarray}
where ${\bf r}_{yq}$ denotes the $q$-th column of ${\bf R}_{\bf y}$, and ${\bf e}_q$ represents the unit column vector with $e_{ql}=\delta(q-l)$ as the $l$-th entry. Next, we define
\begin{eqnarray}
{\bf D} &\!\!\!\!=\!\!\!\!& [\textrm{vec}({\bf p}_1{\bf p}_1^H)~~ \textrm{vec}({\bf p}_2{\bf p}_2^H)~~\cdots ~~\textrm{vec}({\bf p}_K{\bf p}_K^H)]\\
&\!\!\!\!=\!\!\!\!&[{\bf p}_1^*\otimes {\bf p}_1~~{\bf p}_2^*\otimes {\bf p}_2~~\cdots ~~{\bf p}_K^*\otimes {\bf p}_K]_{Q^2\times K}\label{eqn:defineD}\\
&\!\!\!\!\triangleq \!\!\!\!& [{\bf d}_1~{\bf d}_2~~\cdots ~~{\bf d}_K]
\end{eqnarray}
where ``$\otimes$" is the Kronecker product operator, and its $k^{th}$ column is ${\bf d}_k={\bf p}_k^*\otimes {\bf p}_k$. Then (\ref{eqn:theo_cov2}) can be rewritten as
\begin{eqnarray}
{\bf r}\triangleq {\bf r}_y-{\bf r}_w={\bf D}{\bf g}\label{eqn:linear_sys}
\end{eqnarray}
where ${\bf g}=\textrm{diag}({\bf G})=[g_1,\cdots,g_K]^T$. Note that ${\bf r}$ is obtained from ${\bf R}_{\bf y}$ and the noise power $\sigma_w^2$ only, and ${\bf D}$ depends on ${\bf P}$ only. Thus, given ${\bf R}_{\bf y}$, $\sigma_w^2$ and ${\bf P}$, we can obtain a {\em unique} ${\bf g}$ satisfying the $Q^2$ linear equations in (\ref{eqn:linear_sys}), as long as the fat $Q^2\times K$ matrix ${\bf D}$ (due to $Q^2\leq K$) has full rank $K$. As shown in Appendix \ref{sec:matrix_proof}, since $\textrm{rank}({\bf D})=K$, the solution is
\begin{eqnarray}
{\bf g}={\bf D}^{\dag}{\bf r}.\label{eqn:linear_sol}
\end{eqnarray}
Therefore, we complete the proof of Theorem 1. \hfill\QED\

\begin{remark}
Theorem 1 shows that $Q$ pilot REs allow the BS to estimate the large-scale channel gains of up to $Q^2$ users. The key insight behind this fact is that a complex covariance matrix has $Q^2$ degrees of freedom, so that we can collect up to $Q^2$ linearly independent equations.
\end{remark}

In the finite $M$ case where $\Rmh_{\bf y}$  differs from  ${\bf R}_{\bf y}$, the proof of Theorem 1 suggests a method for estimating ${\bf g}$ from the sample covariance $\Rmh_{\bf y}$ in  (\ref{eqn:Ryh}). Following the same vectorization operation as in (\ref{eqn:reshapeR}), we first compute
\begin{equation}
\widehat{\bf r}= \textrm{vec}(\Rmh_{\bf y})-{\bf r}_w,
\end{equation}
and then obtain $\widehat{\bf g}$ (the estimate of ${\bf g}$) as the solution to the following optimization problem:
\begin{subequations}\label{eqn:opt}
\begin{align}
 \minimize_{\thetav}
  & \;\;\; \|\widehat{\bf r}-{\bf D}{\bf \thetav}\|^2 \\
 \textrm{subject to} & \;\;\; {\bf \thetav}\succeq {\bf 0}.
 \end{align}
\end{subequations}
Problem (\ref{eqn:opt}) is a non-negative least-squares (NNLS) problem, a special quadratic programming problem, which has been intensively studied recently \cite{NNLS, NNLS2}, and can be readily solved by general-purpose quadratic-programming solvers\footnote{In fact, seeking efficient and special-purpose solvers of (\ref{eqn:opt}) which exploit the sparsity promoting properties of NNLS are a topic and worth further investigation.}.

It is worth making a few remarks. First, massive arrays improve the estimation performance, as increasing the size of the BS antenna array  improves the sample covariance estimate and thus the quality of the large-scale gain estimates provided by (\ref{eqn:opt}). In addition, the pilot design matrix ${\bf P}$'s, with random entries sufficing in principle systematic designs as we will show later in Sec. \ref{sec:sim}, can provide ${\bf D}$'s with desirable properties and superior performance.

While Theorem 1 is stated for the entries of ${\bf P}$ being complex and generic, we also consider practical constraints such as restricting all pilot values to be real-valued, or have a constant amplitude but random phases. Based on Theorem 1, we have the following two corollaries.

{\bf Corollary 1:} Consider ${\bf P}=(p_{qk})_{Q\times K}$ where $p_{qk}$'s are real and generic. When $M\rightarrow \infty$, the BS is able to identify all the $K$ users almost surely if $K\leq Q(Q+1)/2$.

{\it Proof:} The proof is deferred to Appendix \ref{sec:corollary1_proof}. \hfill\QED

{\bf Corollary 2:} Consider ${\bf P}=(p_{qk})_{Q\times K}$ where $p_{qk}=e^{\theta_{qk}}$ and $\theta_{qk}\sim \mathcal{U}[0,2\pi)$ is i.i.d. generated. When $M\rightarrow \infty$, the BS is able to identify all the $K$ users almost surely if $K\leq Q^2-Q+1$.

{\it Proof:} The proof is deferred to Appendix \ref{sec:corollary2_proof}. \hfill\QED

\begin{remark}
The intuition behind the corollaries above is that the degrees of freedom of their covariance matrices reduce to $Q(Q+1)/2$ and $Q^2-Q+1$, respectively.
\end{remark}

In fact, in both motivating examples we described at the onset, only a fraction of users are expected to have significantly nonzero (or appreciable) large-scale channel gains with respect to the BS, say $K'$ users. If $K'\ll K$, then (\ref{eqn:opt}) becomes a compressed sensing problem, such as in \cite{Sayeed}.

Finally, we briefly discuss the setting with $K>Q^2$ where Theorem 1 cannot be applied. As a matter of fact, as long as $K'\leq Q^2$, certain accurate solutions can still be found with high probability even with $K> Q^2$. In this case, (\ref{eqn:opt}) can be directly solved  using an NNLS solver (without the need for regularization). If $K'\ll K$, then the solution can still be uniquely identified with high probability (see \cite{NNLS}).

\section{Performance Metric Analysis}\label{sec:performance}

In this section, we investigate the estimation performance of large-scale gains. Using ``$\circ$" as the Hadamard product operator and ``$|\cdot|$" as the element-wise amplitude-taking operator, we first show a key property between the column inner products of ${\bf D}$ and ${\bf P}$ in the following proposition:

{\bf Proposition 1:} ${\bf d}_i^H {\bf d}_j=|{\bf p}_i^H{\bf p}_j|^2$ for $\forall i$ and $\forall j$. 
In addition, ${\bf D}^H{\bf D}=({\bf P}^H{\bf P})\circ({\bf P}^H{\bf P})^H=|{\bf P}^H{\bf P}|^{\circ 2}$.

{\it Proof:} We first prove the first equality in the following:
\begin{subequations}
\label{eqn:ddpp}
\begin{eqnarray}
{\bf d}_i^H {\bf d}_j=\sum_{l=1}^{Q^2} d_{li}^*d_{lj}&\!\!\!\!=\!\!\!\!&\sum_{q=1}^Q \Bigg(\sum_{q'=1}^Q (p_{qi}^*p_{q'i})^*(p_{qj}^*p_{q'j})\Bigg)\\
&\!\!\!\!=\!\!\!\!&\sum_{q=1}^Q p_{qi}(p_{qj})^*\!\Bigg(\!\!\sum_{q'=1}^Q p_{q'i}(p_{q'j})^*\!\!\Bigg)^{\!*}\ \ \ \ \\
&\!\!\!\!=\!\!\!\!&|{\bf p}_i^H{\bf p}_j|^2.
\end{eqnarray}
\end{subequations}
It can be seen that no matter if ${\bf D}$ is real-valued or complex-valued, the inner product of its any two columns is always real-valued and non-negative. From (\ref{eqn:ddpp}), we can directly obtain the other equations $({\bf D}^H{\bf D})_{i,j}={\bf d}_i^H {\bf d}_j=|{\bf p}_i^H{\bf p}_j|^2 =|{\bf P}_{:,i}^H{\bf P}_{:,j}|^{\circ 2}=(|{\bf P}^H{\bf P}|^{\circ 2})_{i,j}$ for any $i,j$. \hfill\QED

\begin{remark}
The proposition reveals that ${\bf D}^H{\bf D}$ is real, symmetric, and non-negative, and each of its diagonal entries dominates all the entries on the same row and column.
\end{remark}

In the rest of this section, we restrict our attention to the case of $K\leq Q^2$, i.e., in the range where Theorem 1 is valid. Let us go back to problem (\ref{eqn:opt}) where we aim to minimize $\|\widehat{\bf r}-{\bf D}\thetav\|^2$ subject to $\thetav \succeq {\bf 0}$. Recall that in the finite $M$ case $\widehat{\bf r}$ differ from ${\bf r}$. Letting  ${\bf r}_e = \widehat{\bf r}- {\bf r}$ denote the large-scale gain estimation error and using (\ref{eqn:linear_sys}), we  obtain  
\begin{eqnarray}
\widehat{\bf r}={\bf D}{\bf g}+{\bf r}_e.\label{eqn:perf_linear_sys}
\end{eqnarray}
We ignore the non-negativity constraint on $\{g_k\}$ and thereby focus on the analysis of the ZF estimator:
\begin{eqnarray}
\widehat{\bf g}={\bf D}^{\dag}\widehat{\bf r}=({\bf D}^H{\bf D})^{-1}{\bf D}^H\widehat{\bf r}={\bf g}+\underbrace{({\bf D}^H{\bf D})^{-1}{\bf D}^H{\bf r}_e}_{\triangleq ~{\bf z}:~\textrm{estimation noise}}.
\end{eqnarray}
To see the estimation performance of $\widehat{\bf g}$, we can analyze the noise enhancement, captured by the $K\times K$ covariance matrix of the estimation noise (error) vector ${\bf z}$:
\begin{eqnarray}
{\bf R}_z\triangleq \mathbb{E}({\bf z}{\bf z}^H)&\!\!\!\!=\!\!\!\!&({\bf D}^H{\bf D})^{-1}{\bf D}^H\mathbb{E}({\bf r}_e{\bf r}_e^H){\bf D}({\bf D}^H{\bf D})^{-1}.\ \ \
\end{eqnarray}
As a closed-formed expression of $\mathbb{E}({\bf r}_e{\bf r}_e^H)$ is cumbersome and ${\bf g}$ dependent, we opt to use $\mathcal{CN}({\bf 0},\sigma_e^2{\bf I})$ in place of the true distribution\footnote{It is worth noting that the covariance matrix of ${\bf r}_e$ involves fourth-order statistics of the ${\bf h}_k$'s and the unknown $g_k$'s.  The i.i.d. complex Gaussian assumption represents a worst-case scenario distribution in the context of power constraints,  which  also simplifies performance analysis.} of ${\bf r}_e$, and $\sigma_e^2=\max_q((\mathbb{E}({\bf r}_e{\bf r}_e^H))_{q,q})$. Also, note that this distribution has the same asymptotic behavior as the true distribution, since  $\sigma_e^2\rightarrow 0$ when $M\rightarrow \infty$. Hence, we simply the expression of ${\bf R}_z$ to
\begin{eqnarray}
{\bf R}_z=\sigma_e^2({\bf D}^H{\bf D})^{-1}.
\end{eqnarray}
Inspection of the expression above reveals that ${\bf R}_z$ arises as the combined effect of  $({\bf D}^H{\bf D})^{-1}$  which clearly is a function of   the pilot code design, ${\bf P}$, and of $\sigma_e^2$, which captures the accuracy of the sample covariance estimate and does \emph{not} depend on ${\bf P}$.  As a result, we opt to evaluate the impact in estimation performance of any given pilot code design, via the distribution of the eigenvalues\footnote{Characterizing the distribution of its eigenvalues is still an active research direction in random matrix theory, such as \cite{random_matrix}.} of $ ({\bf D}^H{\bf D})^{-1} $  and their aggregate effect via the trace of $({\bf D}^H{\bf D})^{-1} $. Given that the trace is minimized when  ${\bf D}$ is  unitary, we can view the trace of $({\bf D}^H{\bf D})^{-1} $ as a measure of noice enhancement.

\subsection{Noise Enhancement Analysis}

In this section, we will consider Grassmannian line packing \cite{grassmannian} to design the pilot code sequences, since it provides good properties of the inner product between arbitrary two vectors. Using algorithms such as in \cite{love_grassmannian}, we can find out a $Q\times K$ matrix ${\bf P}$, i.e., we can pack $K$ 1-dimensional lines in the $Q$-dimensional space, so that for any two columns ${\bf p}_i$ and ${\bf p}_j$, the amplitude of their inner product $c_p\triangleq \max_{\bf P}\min_{i\neq j}(|{\bf p}_i^H{\bf p}_j|)$ is as large as possible. Given $Q$ and $K$, \cite{Hearth} (Theorem 2.3 and Corollary 2.4) provided the following lower bound on $c_p$, also referred to as the Welch bound:
\begin{eqnarray}
c_p\geq \sqrt{\frac{K-Q}{Q(K-1)}}.\label{eqn:grassmannian_eq}
\end{eqnarray}
It was shown in \cite{Hearth} that this lower bound is tight and thus $|{\bf p}_i^H{\bf p}_j|=\sqrt{\frac{K-Q}{Q(K-1)}}$ for $\forall i$ and $\forall j\neq i$, when $K\leq Q^2$ if ${\bf P}\in\mathbb{C}^{Q\times K}$, and when $K\leq Q(Q+1)/2$ if ${\bf P}\in\mathbb{R}^{Q\times K}$, i.e., satisfying the condition called the \emph{full frame}. Note that the conditions $K\leq Q^2$ and $K\leq Q(Q+1)/2$ are consistent with our Theorem 1 and Corollary 1 in Sec. \ref{sec:scheme_large}, respectively.

Next, we consider ${\bf P}\in\mathbb{C}^{Q\times K}$, $K\leq Q^2$ and for any two columns of ${\bf P}$ their $c_p=\sqrt{\frac{K-Q}{Q(K-1)}}$ as an example. Because of Proposition 1, denoting by $c_d=c_p^2$, we can explicitly obtain
\begin{eqnarray}
{\bf D}^H{\bf D}=\left[\begin{array}{cccc}1&c_d&\cdots&c_d\\c_d&1&\cdots&c_d\\ \vdots&\vdots&\ddots&\vdots\\c_d&c_d&\cdots&1\end{array}\right]_{K\times K}
\end{eqnarray}
which can be further written into the following compact form:
\begin{eqnarray}
{\bf D}^H{\bf D}=c_d{\bf 1}_{K\times 1}({\bf 1}_{K\times 1})^T+(1-c_d){\bf I}_K,
\end{eqnarray}
where the diagonal entries are 1, and the off-diagonal entries are $c_d$. Hence, its eigenvalues are simply given by:
\begin{eqnarray}
\lambda_k=\left\{\begin{array}{lll}1+(K-1)c_d,&&k=1,\\1-c_d,&&k=2,3,\cdots,K.\end{array}\right.
\end{eqnarray}
Finally, we obtain the eigenvalues of $({\bf D}^H{\bf D})^{-1}$ as follows:
\begin{eqnarray}
\lambda'_k=\left\{\begin{array}{lll}\frac{1}{1+(K-1)c_d},&&k=1,\\\frac{1}{1-c_d},&&k=2,\cdots,K.\end{array}\right.\label{eqn:noise_enhan_coeff}
\end{eqnarray}

\section{Simulation Results}\label{sec:sim}

In this section, we will investigate the noise enhancement via simulation for ${\bf P}\in\mathbb{C}^{Q\times K}$ and $K\leq Q^2$.


First, we consider $K=Q^2$ by employing the codebooks generated by Gaussian complex random variables, the theoretical Grassmannian line packing, and the packet of Grassmannian line packing provided in \cite{grassmannian}, respectively. In particular, we will look into the eigenvalues of corresponding $({\bf D}^H{\bf D})^{-1}$.

\emph{1) Gaussian Random Codebook:} The matrix ${\bf P}$ is constructed by first drawing each entry from $\mathcal{CN}(0,1)$ and then normalizing every column to have unit norm. Once $({\bf D}^H{\bf D})^{-1}$ is obtained, the average noise enhancement per dimension is given by $10\log_{10}(\sum_{k=1}^K \lambda_k'/Q^2)$ dB.

\emph{2) Theoretical Grassmannian-Line-Packing Codebook:} The eigenvalues of $({\bf D}^H{\bf D})^{-1}$ under $K=Q^2$ can be simplified to: $\lambda'_1=1/Q$ and $\lambda'_k=1+1/Q$ for $k=2,\cdots,K$. Thus, the average noise enhancement per dimension is $10\log_{10}(1+(Q-1)/Q^2)$ dB. Note that ${\bf P}$ archiving the equality in (\ref{eqn:grassmannian_eq}) can be explicitly expressed in the closed form for special cases only (see discussion in \cite{Hearth}). In addition, when $Q$ is large, the average noise enhancement per dimension is approximately $(10\log_{10}e)/Q$ dB (close to 0 dB), where $e$ is the natural logarithmic base number.

\emph{3) Simulated Grassmannian-Line-Packing Codebook:} Owing to the existence of the Grassmannian line packing limit but the lack of general close-formed construction, a vast amount of iterative algorithms have been developed to construct a set of vectors approximating the theoretical limit. By running the package provided by \cite{grassmannian} to generate ${\bf P}$ and then $({\bf D}^H{\bf D})^{-1}$, the average noise enhancement per dimension is given by $10\log_{10}(\sum_{k=1}^K \lambda_k'/Q^2)$ dB.

\begin{figure}[!b] \centering \vspace{-0.15in}
\includegraphics[width=2.9in]{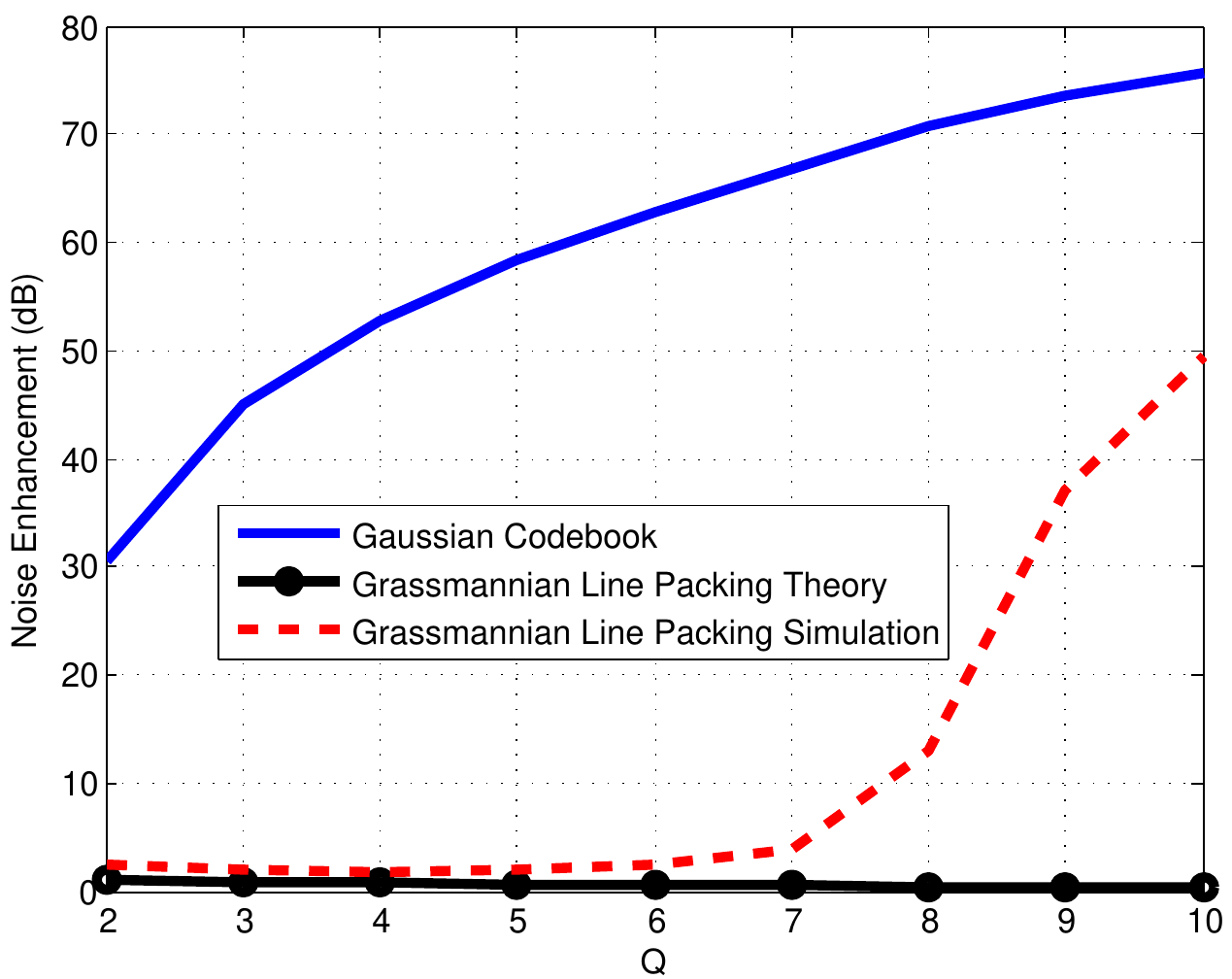}\vspace{-0.15in}
\caption{Comparison of Average Noise Enhancement per Dimension}\label{fig:performanceQ}\vspace{-0.05in}
\end{figure}

\begin{figure}[!b] \centering \vspace{-0.15in}
\includegraphics[width=2.9in]{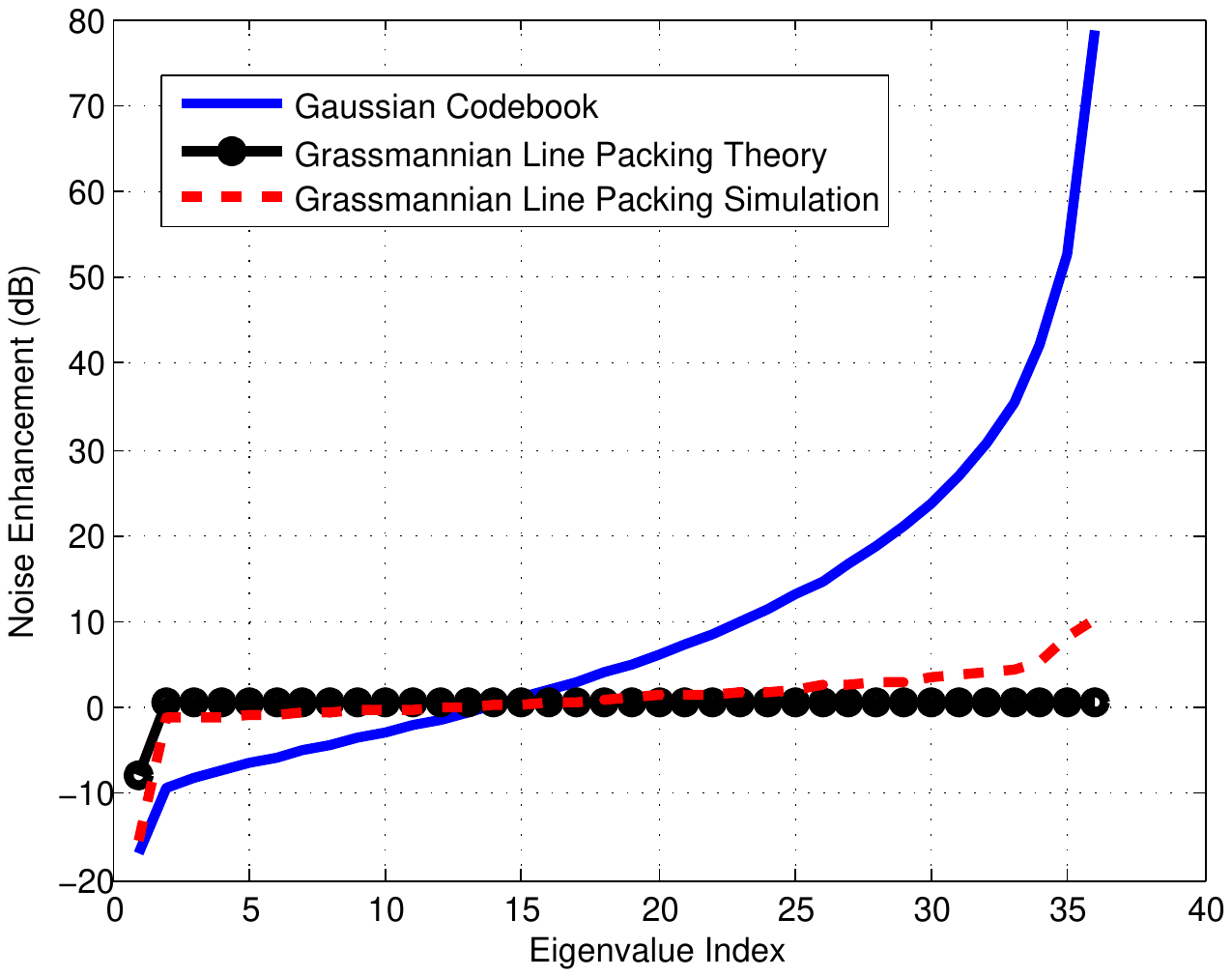}\vspace{-0.15in}
\caption{Comparison of Noise Enhancement of Each Dimension}\label{fig:performance_eigdistribution_Q6}
\end{figure}

Based on the three types of codebook design introduced above, we will show their noise enhancement performance for $K=Q^2$ where $Q=2,3,\cdots,10$. Fig. \ref{fig:performanceQ} shows the comparison of average noise enhancement per dimension. The noise enhancement of the Gaussian codebook could be even 50 dB higher than the Grassmannian approach at $Q=10$, which means that Grassmannian approach is much better. Note that the red dashed curve (line packing algorithms) diverges from the black curve (the theoretical limit) when $Q\geq 7$. This is because when we ran the package provided by \cite{grassmannian}, we kept the output once we were asked whether to stop, which implies that when $Q$ is large, the number of iterations is not large enough to produce more precise result. As we tested for several examples, running more iterations would make the simulated curve closer to the theoretical curve at cost of the time complexity\footnote{It is worth noting that while \cite{Hearth} implies the theoretical Grassmannian line packing limit is achievable, the simulated curve differs from the theoretical curve due to precision allowance and numerical errors accumulated through algorithm iterations. The simulated curve can be made closer to the theoretical one by either running more iterations of existing algorithms or inventing more efficient algorithms, which are not the main focus of this paper.}. In Fig. \ref{fig:performance_eigdistribution_Q6}, we show the the noise enhancement of codebooks in each of the corresponding $Q^2=36$ dimensions. It can be seen that the noise enhancement of the Grassmannian Line Packing codebook is very close to 0 dB in all dimensions. In contrast, although the noise enhancement of the Gaussian codebook is even less in some dimensions, it could much higher (up to 78 dB) in the other dimensions than the Grassmannian approach.

\begin{figure}[!t] \centering 
\includegraphics[width=3.0in]{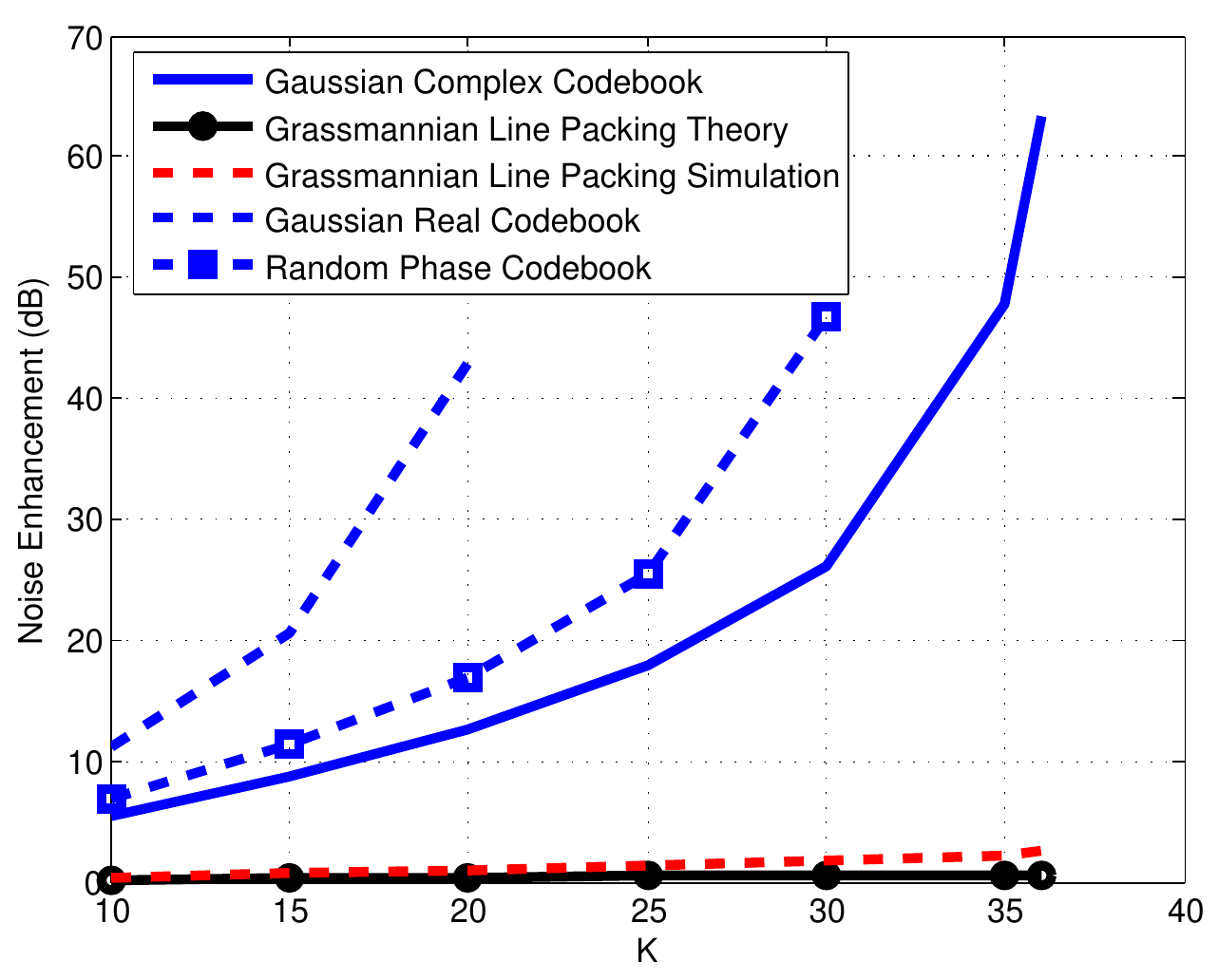}\vspace{-0.15in}
\caption{Comparison of Noise Enhancement of Each Dimension}\label{fig:performanceQK6_all}\vspace{-0.15in}
\end{figure}

Besides $K=Q^2$, we also study the setting with $K<Q^2$ since we might not need to deal with fully loaded systems. Since $K\leq Q$ implies orthogonal pilot design, we consider $Q<K<Q^2$ only. 
Note that for ``Theoretical Grassmannian Line Packing", we need to replace with the following result:
\begin{eqnarray}
10\log_{10}\Big(\frac{Q}{K^2}+\frac{Q(K-1)^2}{(Q-1)K^2}\Big)~~\textrm{dB}.
\end{eqnarray}
Moreover, we also consider the Gaussian real codebook when $K\leq Q^2-Q+1$ and the random phase codebook when $K\leq Q(Q+1)/2$ specified in Sec. \ref{sec:scheme_large}. Fig. \ref{fig:performanceQK6_all} shows the noise enhancement in each dimension in response to $K$ by using the 5 methods of codebook construction for $Q=6$. Clearly, the Grassmannian codebook outperforms all the others.

\section{Conclusion}

In this paper, we propose a novel scheme for estimating large-scale gains of the channels between the users and the BSs on the fly when we apply non-orthogonal pilot codes for UL training in TDD/reciprocity-based systems with massive antenna arrays at the BSs. With $Q$ REs allocated for UL training, the new scheme enables the BS to estimate the large-scale channel gains of up to all $K$ users as long as $K\leq Q^2$. The key of the new scheme is the use of the massive antenna array incorporated with fully exploring the degrees of freedom of the covariance matrix and non-orthogonal pilot codes.

Following our proposed scheme, several interesting problems need further investigation. For example, if channel dependencies exist among the co-located antennas, how to use the sample covariance matrix to better approximate the exact covariance matrix is interesting. For another example, while the Grassmannian line packing codebook is preferable for estimating large-scale gains, it is well know that orthogonal pilot code design is desired for estimating small-scale fading, because projecting the received signal into each spatial dimension does not incur power loss. Therefore, it would be of interest to investigate if there exist any other efficient and intermediate scheme to inherit both of their good features.

\appendix

\subsection{The Proof of $\textrm{rank}(\bf D)=K$ for Theorem 1}\label{sec:matrix_proof}

To show $\textrm{rank}({\bf D})=K$ almost surely for $K\leq Q^2$, it suffices to show $\textrm{det}({\bf D})\neq 0$ almost surely when  $K= Q^2$. Note that $\textrm{det}({\bf D})$ is a polynomial of the generic entries (or variables) $p_{qk}$'s of ${\bf P}$, defined in the continuous field. Thus, $\textrm{det}({\bf D})= 0$ implies that either it always holds for arbitrary choices of $p_{qk}$'s or the finite number of solutions to satisfy $\textrm{det}({\bf D})= 0$ constitute a subset with Lebesgue measure zero. Hence, to prove $\textrm{det}({\bf D})\neq 0$ almost surely, we only need to find one specific choice of $p_{qk}$'s so that $\textrm{det}({\bf D})\neq 0$\footnote{This approach has been widely used in network information theory to study linear independencies among the row/column vectors of a matrix.}. Specifically, we choose ${\bf P}$ to be a Vandermonde matrix:

\small\vspace{-0.15in}
\begin{eqnarray}
{\bf P}=\left[\begin{array}{ccccc}1&a_1&(a_1)^2&\cdots&(a_1)^{Q^2-1}\\
1&a_2&(a_2)^2&\cdots&(a_2)^{Q^2-1}\\
\vdots&\vdots&\vdots&\ddots&\vdots\\
1&a_Q&(a_Q)^2&\cdots&(a_Q)^{Q^2-1}\\\end{array}\right]_{Q\times Q^2}
\end{eqnarray}
\normalsize
where $a_{i}\neq a_j$ for $\forall i$, $\forall j\neq i$. We will show that the resulting ${\bf D}$ is also a Vandermonde matrix. To see this, we write the $k^{th}$ column of ${\bf D}$ for each $k=1,\cdots,Q^2$ as follows:

\small\vspace{-0.15in}
\begin{eqnarray}
&\!\!\!\!\!\!\!\!&{\bf d}_k\triangleq {\bf p}_k^*\otimes {\bf p}_k\notag\\
&\!\!\!\!=\!\!\!\!&[(a_1)^{k-1}\cdots a_Q^{k-1}]^H\otimes [a_1^{k-1}\cdots (a_Q)^{k-1}]^T\\
&\!\!\!\!=\!\!\!\!&\!\![(\!(a_1)^{k\!-1})^* [a_1^{k\!-1}\!\cdots a_Q^{k-1}],\cdots,\!(\!(a_Q)^{k\!-1})^* [a_1^{k-1}\!\cdots a_Q^{k\!-1}\!]]^T\ \ \ \ \\
&\!\!\!\!=\!\!\!\!&\!\![(a_1^*)^{k-1} [a_1^{k-1}\cdots a_Q^{k-1}],~\cdots,~ (a_Q^*)^{k-1} [a_1^{k-1}\cdots a_Q^{k-1}]]^T\\
&\!\!\!\!=\!\!\!\!&[(a_1^*a_1)^{k-1},\cdots, (a_1^*a_Q)^{k-1},(a_2^*a_1)^{k-1},\cdots, (a_2^*a_Q)^{k-1},\notag\\
&\!\!\!\!\!\!\!\!& \cdots,(a_Q^*a_1)^{k-1},\cdots,(a_Q^*a_Q)^{k-1}]^T.
\end{eqnarray}
\normalsize
It can be seen that each entry has the same exponential factor $k-1$ for the $k^{th}$ column of ${\bf D}$. Next, we denote by ${\bf d}_2=[b_1,b_2,\cdots,b_{Q^2}]^T$, and its $l^{th}$ entry can be easily written as
\begin{eqnarray}
b_{l}=a_{l_1}^*a_{l_2},~~~~~~l=1,\cdots,Q^2
\end{eqnarray}
where $l_2=\textrm{mod}(l-1,Q)+1$ and $l_1=(l-l_2)/Q+1$, i.e., $l=l_2+(l_1-1)Q$. Thus, we directly obtain

\small\vspace{-0.1in}
\begin{eqnarray}
\textrm{det}({\bf D})=\prod_{1\leq i<j\leq Q^2}(b_j-b_i)=\prod_{1\leq i<j\leq Q^2}(a_{j_1}^*a_{j_2}-a_{i_1}^*a_{i_2}),
\end{eqnarray}
\normalsize
where $i=i_2+(i_1-1)Q$, $j=j_2+(j_1-1)Q$. Given any $i,j$ and $i\neq j$, we must have $(i_1,i_2)\neq (j_1,j_2)$. Since each $a_q$, $q=1,\cdots,Q$ is generic, we have $b_j-b_i\neq 0$ almost surely, which directly implies $\textrm{det}({\bf D})\neq 0$ almost surely.

\subsection{The Proof of Corollary 1}\label{sec:corollary1_proof}

Observations of (\ref{eqn:theo_cov}) reveal that ${\bf R}_{\bf y}-\sigma_w^2{\bf I}={\bf P}{\bf G}{\bf P}^H={\bf P}{\bf G}{\bf P}^T$ is real symmetric, which means that the entries in its upper triangle are the same as those in the lower triangle. Thus, the number of equations that we need to consider are only those corresponding to the diagonal entries and the entries in one triangle, which is given by $Q+(Q^2-Q)/2=Q(Q+1)/2$. Translating this observation into examining each row of the resulting matrix ${\bf D}$ implies that its $i^{th}$ row and $j^{th}$ row are identical, whenever $i=l_2+(l_1-1)Q$, $j=l_1+(l_2-1)Q$, for $\forall i$, $\forall j\neq i$. If we delete the repeated $Q(Q-1)/2$ rows of ${\bf D}$ and also the last $Q(Q-1)/2$ columns of ${\bf D}$, the resulting $(Q(Q+1)/2)\times (Q(Q+1)/2)$ matrix ${\bf D'}$ is still a Vandermonde matrix. Following the similar proof, we can show ${\bf D'}$ has full rank almost surely. Overall, if $K\leq Q(Q+1)/2$, (\ref{eqn:linear_sys}) are under-constrained, and thus ${\bf g}$ can be uniquely determined.

\subsection{The Proof of Corollary 2}\label{sec:corollary2_proof}

Observations of (\ref{eqn:theo_cov}) reveal that the diagonal entries of ${\bf R}_{\bf y}-\sigma_w^2{\bf I}={\bf P}{\bf G}{\bf P}^H$ is a scaled identity matrix, which means that the $l^{th}$ row of ${\bf D}$ is identical to its first row, when $l=q+(q-1)Q$ for every $q=2,\cdots,Q$. If we delete the repeated $Q-1$ rows of ${\bf D}$ and also the last $Q-1$ columns of ${\bf D}$, the resulting $(Q^2-Q+1)\times (Q^2-Q+1)$ matrix ${\bf D'}$ is again a Vandermonde matrix. Following the similar proof, we can show ${\bf D'}$ has full rank almost surely. Overall, if $K\leq Q^2-Q+1$, (\ref{eqn:linear_sys}) is under-constrained, and ${\bf g}$ can be uniquely determined.


%
%
%
%

\begin{thebibliography}{1}


\bibitem{Marzetta_massive_mimo}
T. L. Marzetta, ``How much training is required for multiuser MIMO?," {\em Proc. 2006 Fortieth Asilomar Conference on Signals, Systems
and Computers,} pp.359 -- 363, Nov. 2006.

\bibitem{Ansuman-JSDM}
A. Adhikary, J. Nam, J. Ahn, and G. Caire, ``Joint Spatial Division and Multiplexing -- The Large-Scale Array Regime," {\em IEEE Transactions on Information Theory,} Vol. 59, Issue 10, pp.6441--6463, Oct. 2013.

\bibitem{Marzetta_noncooperative}
T. Marzetta, ``Noncooperative cellular wireless with unlimited numbers of base station antennas," {\em IEEE Trans. on Wireless Commun.,} vol. 9, no. 11, pp. 3590--3600, Nov. 2010.

\bibitem{Ozgun_ICC2016}
Z. Li, N. Rupasinghe, O. Y. Bursalioglu, C. Wang, H. Papadopoulos, and G. Caire, ``Directional training and fast sector-based processing schemes for mmwave channels," {\em IEEE ICC 2016}, May 2016.

\bibitem{wbe_full}
C. Wang, O. Y. Bursalioglu, H. Papadopoulos, and G. Caire, ``On-the-fly Large-scale Channel-Gain Estimation for Massive Antenna-Array Base Stations," {\em in Prepareation}, 2018.

\bibitem{random_matrix}
S. Kumar, and Z. Ahmed, ``Spectral statistics for ensembles of various real random matrices," {\em arxiv: 1704.02715 [quant-ph]}, April 2017.

\bibitem{love_grassmannian}
D. J. Love, ``Grassmannian Subspace Packing," {\em [Available Online] https://engineering.purdue.edu/\%7Edjlove/grass.html}.

\bibitem{Hearth}
T. Strohmer, and R. W. Heath Jr., ``Grassmannian Frames with Applications to Coding and Communication," 2002, {\em [Available online] https://www.math.ucdavis.edu/\%7Estrohmer/papers/2002/grass.pdf}.

\bibitem{grassmannian}
A. Medra, and T. N. Davidson, ``Flexible codebook design for limited feedback systems via sequential smooth optimization on the Grassmannian manifold," {\em IEEE Trans. Signal Processing,} Vol. 62, Issue 5, pp. 1305--1318, March 2014. The simnulation package is available at {\em \url{http://www.ece.mcmaster.ca/~davidson/pubs/Flexible_codebook_design.html}}.


\bibitem{Sayeed}
W. U. Bajwa, J. Haupt, A. M. Sayeed, and R. Nowak, ``Compressed channel sensing: A new approach to estimating sparse multipath channels," {\em Proc. of the IEEE,} vol. 98, no. 6, pp. 1058-1076, Jun. 2010.

\bibitem{NNLS}
V. Sharan, K. Tai, P. Bailis, and G. Valiant, ``There and Back Again: A General Approach to Learning Sparse Models," {\em arXiv:1706.08146v1 [cs.LG]}, June, 2017.

\bibitem{NNLS2}
S. Foucart, and D. Koslicki, ``Sparse Recovery by Means of Nonnegative Least Squares," {\em IEEE Signal Processing Letters,} Vol. 21, Issue 4, pp.498--502, April 2014.

\end{thebibliography}
\end{document}